\documentclass{llncs}
\usepackage{graphicx}

\begin{document}
\mainmatter              % start of the contributions
\title{Memory and compiler optimizations
for low-power and -energy}
\titlerunning{Compiler optimizations 
for low-energy}  % abbreviated title (for running head)
%                                     also used for the TOC unless
%                                     \toctitle is used
%
\author{Olivier Zendra}
\authorrunning{Olivier Zendra}   % abbreviated author list (for running head)
%
%%%% modified list of authors for the TOC (add the affiliations)
\tocauthor{Oliver Zendra (INRIA - LORIA)}
\institute{INRIA-Lorraine / LORIA, Building C, \\
615 Rue du Jardin Botanique, BP 101, \\
54602 Villers-L\`{e}s-Nancy Cedex,\\
 FRANCE\\
\email{Olivier.Zendra@loria.fr}\\ 
\texttt{http://www.loria.fr/\homedir zendra}
}

\maketitle              % typeset the title of the contribution

\begin{abstract}
Embedded systems become more and more widespread, especially
autonomous ones, and clearly tend to be ubiquitous.
In such systems, low-power and low-energy usage get ever more
crucial. 
Furthermore, these issues also become paramount in (massively)
multi-processors systems, either in one machine or more widely in a
grid.  
The various problems faced pertain to autonomy, power supply
possibilities, thermal dissipation, or even sheer energy cost.

Although it has since long been studied in harware, energy
optimization is more recent in software.
In this paper, we thus aim at raising awareness to low-power and
low-energy  issues in the language and compilation community.
We thus broadly but briefly survey techniques and
solutions to this energy issue, focusing on a few specific aspects in
the context of compiler optimizations and memory management.

\end{abstract}
%
%%%%%%%%%%%%%%%%%%%%%%%%%%%%%%%%%%%%%%%%%%%%%%%%%%%%%%%%%%%%%%%%%%%%%%
\section{Introduction}

Embedded systems become more and more widespread, especially
autonomous ones, and clearly tend to be ubiquitous.
In such systems, low-power and low-energy usage get ever more
crucial. 
Furthermore, these issues also become paramount in (massively)
multi-processors systems, either in one machine or more widely in a
grid.  
The various problems faced pertain to autonomy, power supply
possibilities, thermal dissipation, or even energy cost.

They can be addressed from various viewpoints, either in hardware or
software. 
Among those, hardware design implies work at the microelectronics and
physics level, while hardware optimization incurs on-line logic,
dedicated circuits and thus some overhead at runtime (energy and/or
time). 

Software optimization at compile-time, on the other side, is an
off-line logic (if performed statically) that incurs no runtime overhead.
It can also use from more resources (time, memory...), which allows
larger contexts.  
It may however have trouble capturing the exact runtime behavior of
a system.

Energy considerations are relatively recent in compilation, which
historically focused more on size and speed.  
Optimizations for speed and energy are often related \cite{lee1997a}
but not always.
For example, moving some work out of the critical path is good for
time, but it may simply not impact energy, since the work still has to
be done.  
It is also worth stressing that optimizing for (peak) power is
different from optimizing for energy (average power over time) and
different from optimizing for power density (temperature and hot
spots).

In this paper, we aim at raising awareness to low-power and low-energy
issues in the language and compilation community.
We thus perform a broad but shallow survey of techniques and
solutions, focusing on a few specific aspects, all in the context of
compiler optimizations and memory management.

%%%%%%%%%%%%%%%%%%%%%%%%%%%%%%%%%%%%%%%%%%%%%%%%%%%%%%%%%%%%%%%%%%%%%%
\section{Transitions and commutations}

One of the lowest levels at which compilation can address energy
issues is the bit level.
Transitions (bit commutations) between successive instructions cost
energy. 
A compiler can reschedule instructions to minimize this cost
\cite[p.193]{graybill2002a}.

One example is register renaming in order to decrease commutations for
the register name field of instructions. 
Commutation activity on this field was reduced by 11\% in
\cite{kandemir2000a}. 
The overall, global impact of this specific optimization has
nonetheless to be evaluated at whole program level.

%%%%%%%%%%%%%%%%%%%%%%%%%%%%%%%%%%%%%%%%%%%%%%%%%%%%%%%%%%%%%%%%%%%%%%
\section{Loop optimizations}

Very numerous research works have revolved around loop optimizations. 
These were historically targeted to speed.
Detailing all these works falls beyond the scope and space of this
paper of this paper, but we can mention a few ones. 

One canonical example is loop unrolling, where 1 loop with length $n$
running $i$ times becomes 1 loop with length $n*x$ running $n/x$ times:

\begin{verbatim}
for(i=0;i<10000;i++){                            for(i=0;i<5000;i++){
   a();b();            ==== loop unrolling ===>     a();b();
}                                                   a();b();  }                
\end{verbatim}

The impact of loop unrolling is twofold. 
First, the static instruction count increases, since some instructions
are duplicated. 
This leads to a larger code size, hence an increase in energy usage.
But a second effect is to decrease the number of
dynamic instructions, since less are executed for loop control.
This translates into a gain in time as well as in energy. 
It is thus important to balance overhead and gain when using
loop unrolling in an energy-sensitive context.

More examples of loop-oriented optimization for energy will follow
in the context of memory management.

%
%%%%%%%%%%%%%%%%%%%%%%%%%%%%%%%%%%%%%%%%%%%%%%%%%%%%%%%%%%%%%%%%%%%%%%
\section{Execution modes}

This idea behind program modes is to follow program phases.
The most famous example of execution mode is DVS/DFS (Dynamic Voltage
Scaling / Dynamic Frequency Scaling), which deals with the CPU. 
Since $P = C.V^{2}.f$ and $E = P_{avg}.Time$, DVFS consists in decreasing
voltage and frequency (both are tied) to save (dynamic) power and energy.

Other resources also feature execution modes targeted to low energy,
such as sleep modes, or hibernation.
Using these modes aims at 0 consumption (unused resource).
Both dynamic and static power are concerned, which is very effective.

Hardware is able to easily detect phases of low utilization of one
resource, but only {\it a posteriori}.
It has more difficulties and less certainty when trying to {\it
  predict} future low usage phases.
This incurs useless delay before appropriate action (put into
some sleep mode) is taken, as the following figure shows:.
\vspace{-5mm}
\begin{figure*}[h!]
\centerline{\includegraphics[width=10cm]{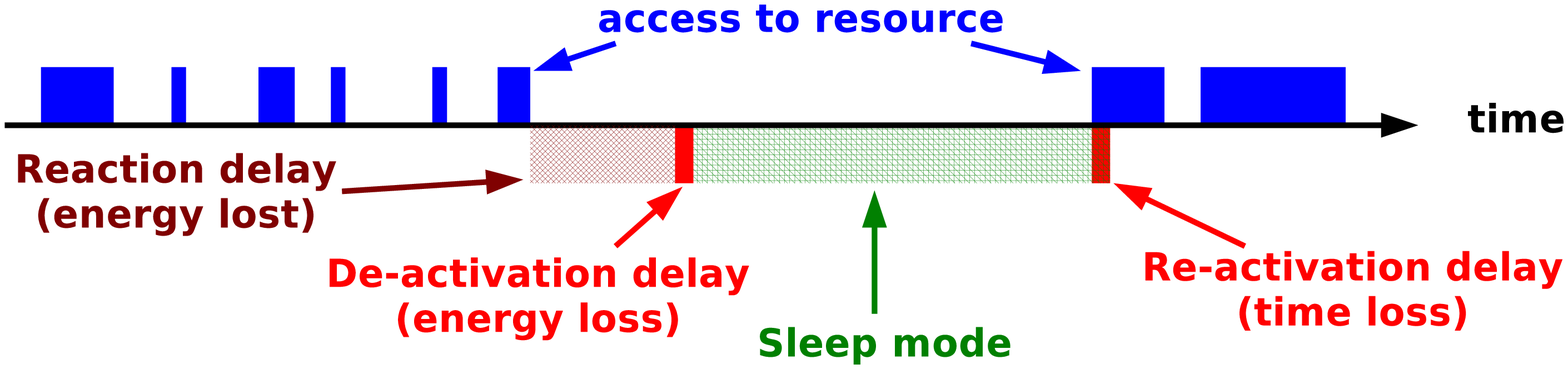}}
\end{figure*}

A compiler, on the contrary, can spot the points where a resource is
(going to be) unused.
The latter thus can be put into sleep mode immediately.
Furthermore, a compiler can even "warn" in advance the hardware of
a future sleep period for a resource, as well as of a future
re-start. 
There is thus no unneeded delay when going to sleep mode or waking-up
a resource, which provides much better results, both in terms of
energy and time:
\vspace{-5mm}
\begin{figure*}[h!]
\centerline{\includegraphics[width=10cm]{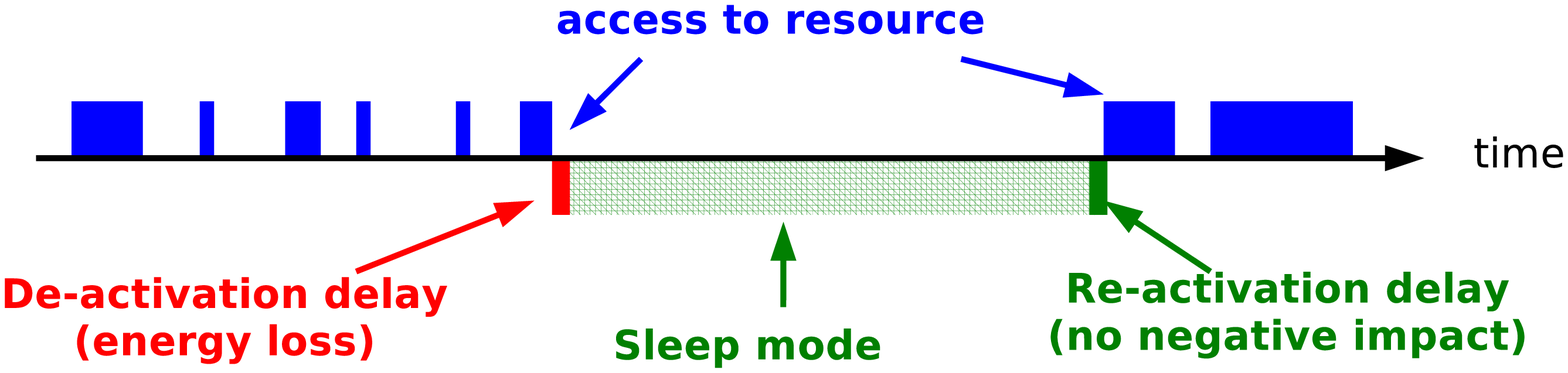}}
\end{figure*}

In the following sections, we will develop on the role of compilation
and memory management to take advantage of sleep modes.

%%%%%%%%%%%%%%%%%%%%%%%%%%%%%%%%%%%%%%%%%%%%%%%%%%%%%%%%%%%%%%%%%%%%%%
\section{Register window}

Register windows consist in having more virtual register than actual
(physical) ones, the virtual registers being separated in several sets
called ``register windows''.
Of course, only 1 register window can be active (used) at a time.

The underlying principle is to change register window according to
program phases: one phase runs into one window.
This makes it possible to reduce register spill (use of main memory
when not enough registers are available), at the cost a some management
overhead (to handle window changes, registers are swapped with
memory).

Working with registers rather than memory means decreasing the number
of transfers and increasing speed  (the very reason register windows
were created for), by up to 11\% according to \cite{ravindran2005a}.

But this also offers some increased opportunities for sleep modes.
In terms of energy, register windows can thus lead to important
saving, up to 25\% \cite{ravindran2005a}.

%%%%%%%%%%%%%%%%%%%%%%%%%%%%%%%%%%%%%%%%%%%%%%%%%%%%%%%%%%%%%%%%%%%%%%
\section{Memory management: compaction}

The idea behind compaction for low energy is very basic: less space
generally means less energy (to hold code or data), and possibly more
opportunities for memory banks sleep modes.
There are many ways to have compact information.

The first one is {\bf contiguity}, which consists in allocating and
keeping data in a minimum of well-filled areas, the others being put
into low-power mode.
To achieve this goal, data moves may be necessary to avoid
fragmentation.
Note that contiguity may be antagonist to speed, since the latter may
benefit from parallel accesses to several memory banks.

{\bf Coalescing} is a way to decrease the space taken by pieces of
data. 
Coalescing variables consists in fitting several ``small'' pieces of
data into only 1 slot.
Subword data and bitwidth aware register allocation are examples of
``spatial'' data coalescing.
\vspace{-5mm}
\begin{figure*}[h!]
\centerline{\includegraphics[width=8cm]{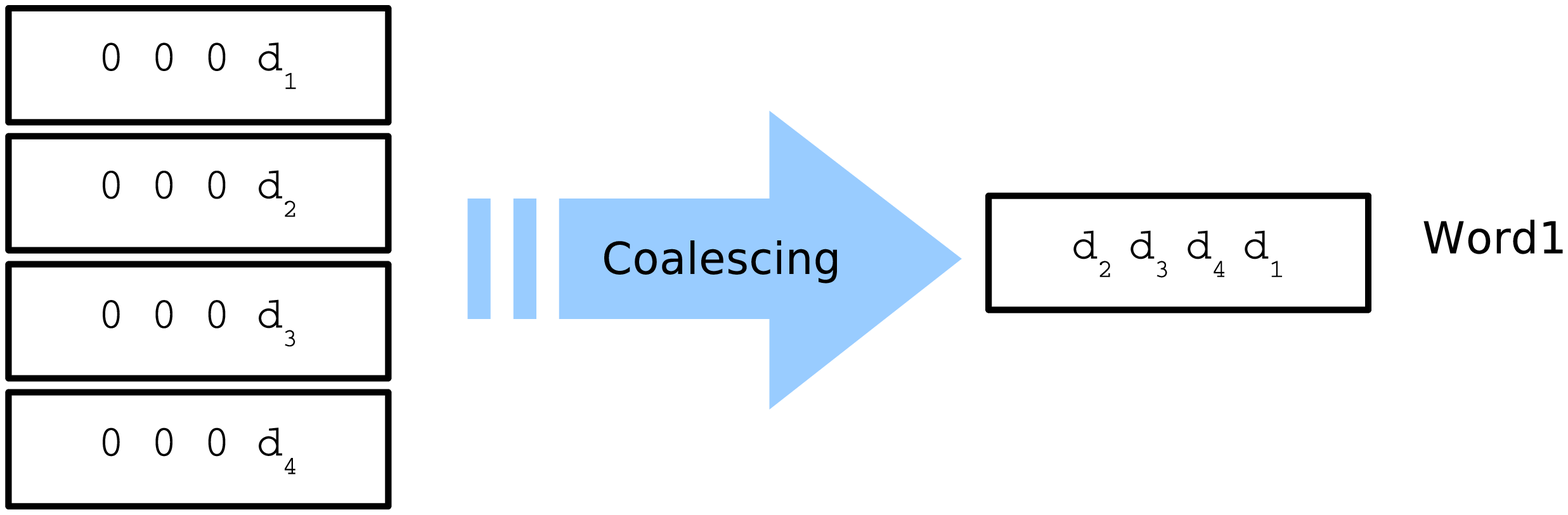}}
\end{figure*}

\vspace{-5mm}
Lifetime analysis is a dual approach allowing ``temporal coalescing''.
It consists in putting in the same slot pieces of data that do not
coexist at the same time, even though each piece completely fills the
slot.

Code or data {\bf compression} is well known and works on a much
larger scale. 
It thus offers stronger opportunities to decrease size (especially for
data), hence helping increase sleep modes usage.
However, the potential overhead of compression is high.
Compression thus appears more adapted to long-lived and seldom
accessed data.

Overall the impact of compaction techniques can be very significant.
On variables,\cite{zhuang2003a} reports 3\% less cycles and 69\%
smaller stack, while \cite{tallam2003a} reaches 10 to 50\% saving in
the number of registers needed.
In \cite{zhang2002a}, data (fields) compression allows 25\% decrease
in heap size and 30\% in energy, while runtime decreases by 12\% and
even 30\% when ISA Data Compression eXtensions are available.

%%%%%%%%%%%%%%%%%%%%%%%%%%%%%%%%%%%%%%%%%%%%%%%%%%%%%%%%%%%%%%%%%%%%%%
\section{Resource access scheduling}

Another very effective way of saving energy in software is to improve
the locality of accesses to resources.
Indeed, grouping accesses increases the length of the
periods over which a specific resource is unused, and can thus be put
into some low-power sleep mode. 

Access re-scheduling occurs at code level, and consist in clustering
 and advancing some accesses:
\vspace{-5mm}
\begin{figure*}[h!]
\centerline{\includegraphics[width=10cm]{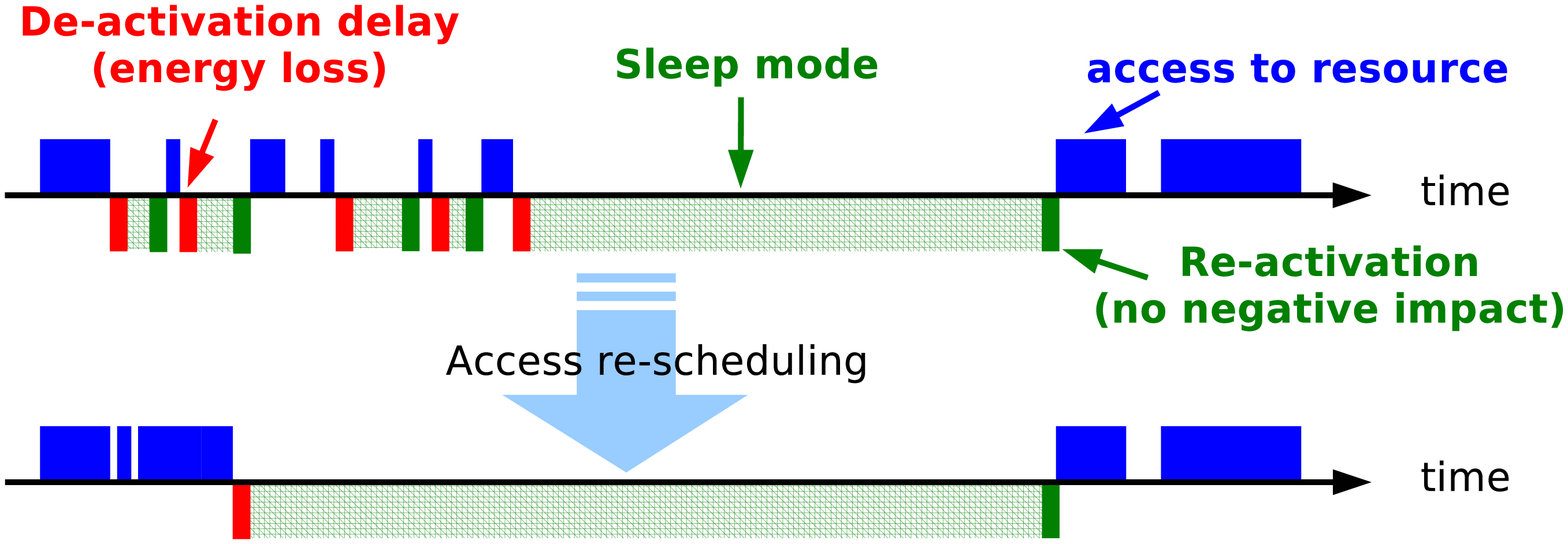}}
\end{figure*}

A specific case of access re-scheduling can be performed at loop level.
Loop fission consists in splitting 1 loop into several loops, when the
instructions in the original loop do not depend on each other.
The various loop can thus process different pieces of data
(arrays...), which offers a better locality and increases sleep mode
opportunities \cite[ch10]{graybill2002a}.

Access re-scheduling may also be performed at the data level, by
changing data layout.
This is dual of code change and is especially interesting for arrays. 
Indeed, accessing those according to layout makes it possible not only
to save time but also to reach up to 10\% saving in energy wrt. 
to basic mode control \cite{athavale2001a}.
Similarly, interlacing arrays that are accessed simultaneously, as
shown below, offers 
8\% savings in energy over  basic mode control \cite{athavale2001a}: 
\vspace{-5mm}
\begin{figure*}[h!]
\centerline{\includegraphics[width=10cm]{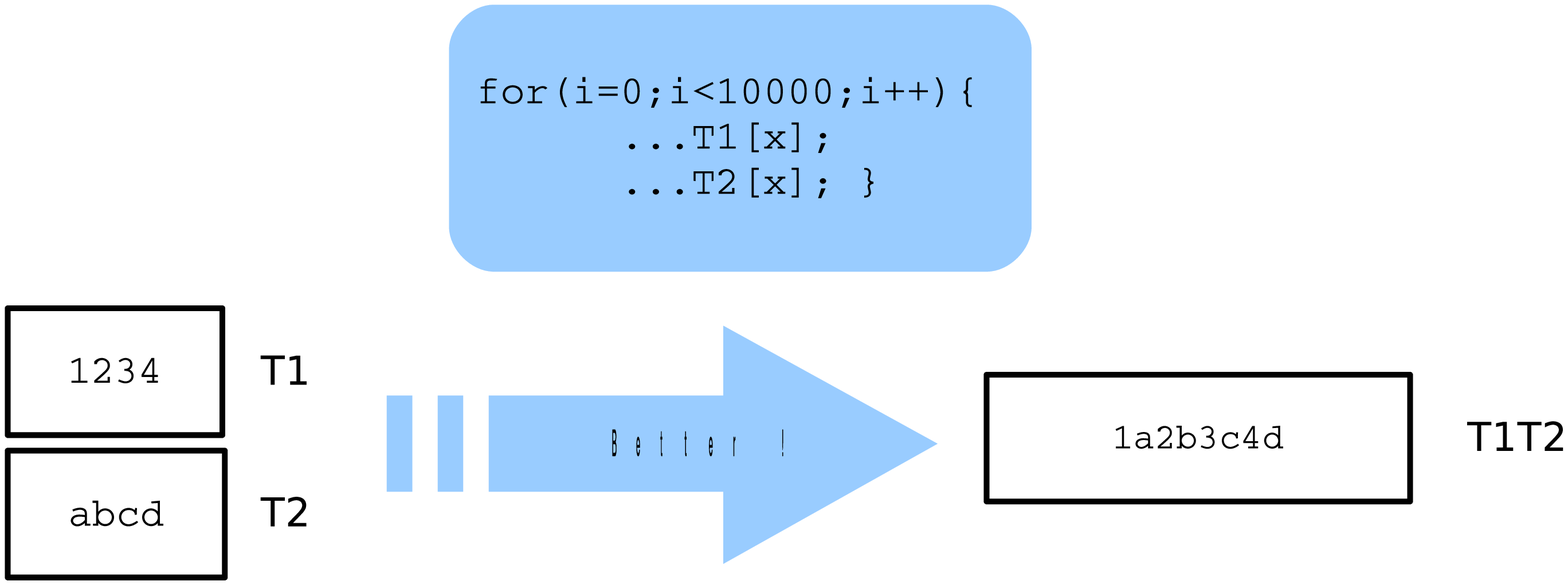}}
\end{figure*}

%%%%%%%%%%%%%%%%%%%%%%%%%%%%%%%%%%%%%%%%%%%%%%%%%%%%%%%%%%%%%%%%%%%%%%
\section{Scratch-pad memory}

Memory-related issues can be of tremendous importance when trying to
save energy.
Some studies even claim that 70\% to 90\% of energy in 2010 will be
used by memory \cite{itrs}.

It is well known that caches generally provide great speed gains.
However, they are poorly adapted to embedded systems, especially
autonomous ones.
They indeed increase the circuit size (for the cache memory as well as
its management logic), are very energy-hungry and are poorly
predictable, which is an issue with real time systems.
Many cacheless systems thus remain in the embedded world.

Scratch-Pad Memory (SPM) is intended to avoid the main drawbacks of
caches.
They consists of small, fast memory areas (SRAM...), very much like
caches, but are directly and explicitly managed at the software level,
either by the developer or by the compiler.
Hence, no dedicated circuit is required for SPM management.

Their advantages compared to caches are numerous \cite{banakar2002a}. 
SPM takes up to 34\% less area than a cache: only the memory is
present, without additional logic.
Their cost is thus lower.
Furthermore, their explicit management makes them more predictable for
real-time systems.
Finally, SPM uses up to 40\% less energy than caches.

\smallskip

SPM can be used throughout a wide range of application domains.
They perform greatly if data accesses are known and regular, which is
for example the case in matrix multiply, audio-video compression
algorithms, filtering... 
They also perform well --- and better than caches --- if the mapping
of data in the SPM is optimal based on access probabilities. 
This has been proven in  \cite{absar2005a}  for structures like lists,
n-trees with low-variation topology

\smallskip

The management of the SPM significantly impacts how it performs.

Static management is a first kind of SPM management strategy.
There, choices (that is data placements) are performed entirely off
line, at compile time, and no data move ever occurs.
These strategies can nonetheless take into account some runtime
information thanks to execution profiles.
They generally feature good performance and good real time
characteristics.

Dynamic strategies are more complex and more recent but have a
tremendous potential.  
There, the  allocation into the SPM is dynamic, thats is performed at
runtime, although it is generally decided at compile time. 
(Dis)placements may occur at runtime.
The main interest of dynamic SPM management techniques lies in the fact
they can take into account regions in the program, or program phases,
instead of considering the program as a whole.
Allocation choices are based on usage frequency, transfer
costs (between SPM and memory), size of objects...

A dynamic management of SPM has many advantages.
It allows better memory (re)use, since freeing SPM immediately is
possible when a piece of data ceases (maybe temporarily) to be in
use. 
It is also better on more complex situations (MPEG21, MPEG4), with 
dynamic creation of tasks, variable data size... 
However, the dynamic management of SPM tends to make real time
constraints harder to meet.
Furthermore, it increases a bit the size of the program, because of the
(software) logic needed. 
Dynamic SPM management also incurs an overhead, both in terms of
energy and time, compared to static SPM management.
The dynamic logic is indeed more complex, and transfers between SPM and
RAM are expensive.
Note that this cost can to some extend be decreased with DMA support
\cite{poletti2004a}, or eliminated completely when direct allocation
in SPM (without coming first from RAM) is possible.

Overall, dynamic management of SPM results in 35\% runtime and 40\%
energy savings wrt. a static placement (except heap) in SPM.

%%%%%%%%%%%%%%%%%%%%%%%%%%%%%%%%%%%%%%%%%%%%%%%%%%%%%%%%%%%%%%%%%%%%%%
\section{Global system analysis}

It is well-known in compilation that system-wide analysis makes it
possible to perform more aggressive and precise (inter-program)
optimizations. 
In the specific context of energy, this holds true as well.

%%% XXXXXXX BEGIN REMOVE IF SPACE NEEDED:
%%Inter-program optimization is important, but is not always easy. 
%%Working at the scheduling level is intrinsically inter-program.
%%Similarly, optimizations performed in hardware consider all programs,
%%but as they are executed, not as a whole.
%%Memory management optimizations performed at the applications level
%%remain single-program, while those performed at OS level are
%%multi-program oriented.
%%In compilation, optimizations are often mono-program, especially with
%%static compilation, that is often separate compilation.
%%Dynamic compilation (JVMs...) however considers  multi-program.
%%% XXXXXXX END REMOVE IF SPACE NEEDED

%%It is overall important to remember that global system analysis is
%%crucial to maximize gains. %%% REMOVE IF SPACE...XXXXXX
As an example, \cite{hom2005a} considered buffer sizing and access
clustering.
They were able to reduce energy usage by 7\% to 49\% with
multi-program optimization wrt. mono-program optimization.

%%%%%%%%%%%%%%%%%%%%%%%%%%%%%%%%%%%%%%%%%%%%%%%%%%%%%%%%%%%%%%%%%%%%%%
\section{Conclusion and perspectives}

Hardware and compilation complete each other.
Compilation can take into account a much larger context, since it has
lots of resources (especially static compilation).
But it may not catch the very exact runtime behavior of programs,
unlike hardware.
Trying to have the best of both worlds is tenting: optimizing Virtual
Machines do it. 
But they remain expensive in terms of resources at runtime.

We think there is thus a clear need of support for hardware-software
(compiler) interface at the ISA level, in order to increase the
synergies between the two worlds.
This way, ``direct'' management of resources by the compiler would be
possible, as well as co-optimizations involving compiler and hardware,
with information transmissions both ways.

VLIW processors and the EPIC architecture offer a higher potential of 
parallelism.
This would of course increase speed, be can also be of interest
energy-wise.
The problem with these architectures is that the compiler has to
provide the parallelism, with requires a lot of work not yet done for
generic processors.

Finally, we  mentioned the paramount importance of memory
for energy usage optimization.
We want once again to stress the
immediate potential of SPM (scratch-pad memory), as well as the ---
possibly longer-term --- potential of Energy-aware Garbage Collectors.

%
% ---- Bibliography ----
%

\nocite{*} 

\bibliographystyle{plain}

\bibliography{zendra_icooolps_2006}

\begin{thebibliography}{10}

\bibitem{absar2005a}
Mohammed~Javed Absar and Francky Catthoor.
\newblock Compiler-based approach for exploiting scratch-pad in presence of
  irregular array access.
\newblock In {\em DATE}, pages 1162--1167, 2005.

\bibitem{athavale2001a}
R.~Athavale, Narayanan Vijaykrishnan, Mahmut~T. Kandemir, and Mary~Jane Irwin.
\newblock Influence of array allocation mechanisms on memory system energy.
\newblock In {\em IPDPS}, page~3, 2001.

\bibitem{avissar2002a}
Oren Avissar, Rajeev Barua, and Dave Stewart.
\newblock An optimal memory allocation scheme for scratch-pad-based embedded
  systems.
\newblock {\em Transaction. on Embedded Computing Systems.}, 1(1):6--26, 2002.

\bibitem{banakar2002a}
Rajeshwari Banakar, Stefan Steinke, Bo-Sik Lee, M.~Balakrishnan, and Peter
  Marwedel.
\newblock Scratchpad memory: design alternative for cache on-chip memory in
  embedded systems.
\newblock In {\em 10th international symposium on Hardware/software codesign
  (CODES'02)}, pages 73--78, New York, NY, USA, 2002. ACM Press.

\bibitem{delaluz2002a}
V.~Delaluz, M.~Kandemir, N.~Vijaykrishnan, M.~J. Irwin, A.~Sivasubramaniam, and
  I.~Kolcu.
\newblock Compiler-directed array interleaving for reducing energy in
  multi-bank memories.
\newblock In {\em 2002 conference on Asia South Pacific design automation/VLSI
  Design (ASP-DAC'02)}, page 288, Washington, DC, USA, 2002. IEEE Computer
  Society.

\bibitem{dominguez2005a}
Angel Dominguez, Sumesh Udayakumaran, and Rajeev Barua.
\newblock Heap data allocation to scratch-pad memory in embedded systems.
\newblock {\em Journal of Embedded Computing (JEC)}, 1(4), 2005.

\bibitem{graybill2002a}
Robert Graybill and Rami Melhem.
\newblock {\em Power aware computing}.
\newblock Kluwer Academic Publishers, Norwell, MA, USA, 2002.

\bibitem{hom2001a}
J.~Hom and U.~Kremer.
\newblock Energy management of virtual memory on diskless devices.
\newblock In {\em Workshop on Compilers and Operating Systems for Low Power
  (COLP'01)}, Barcelone, Espagne, September 2001.

\bibitem{hom2005a}
Jerry Hom and Ulrich Kremer.
\newblock Inter-program optimizations for conserving disk energy.
\newblock In {\em 2005 international symposium on Low power electronics and
  design (ISLPED'05)}, pages 335--338, New York, NY, USA, 2005. ACM Press.

\bibitem{itrs}
ITRS.
\newblock International technology roadmap for semiconductors, 2005.
\newblock http://public.itrs.net.

\bibitem{kandemir2000a}
M.~Kandemir, N.~Vijaykrishnan, M.J. Irwin, W.~Ye, and I.~Demirkiran.
\newblock Register relabeling: A post compilation technique for energy
  reduction.
\newblock In {\em Workshop on Compilers and Operating Systems for Low Power
  (COLP'00)}, Philadelphie, PA, USA, October 2000.

\bibitem{lee1997a}
M.~Lee, V.~Tiwari, S.~Malik, and M.~Fujita.
\newblock Power analysis and minimization techniques for embedded dsp software.
\newblock {\em IEEE Transactions on Very Large Scale Integration}, 5, March
  1997.

\bibitem{poletti2004a}
Francesco Poletti, Paul Marchal, David Atienza, Luca Benini, Francky Catthoor,
  and Jose~Manuel Mendias.
\newblock An integrated hardware/software approach for run-time scratchpad
  management.
\newblock In {\em DAC}, pages 238--243, 2004.

\bibitem{ravindran2005a}
Rajiv~A. Ravindran, Robert~M. Senger, Eric~D. Marsman, Ganesh~S. Dasika,
  Matthew~R. Guthaus, Scott~A. Mahlke, and Richard~B. Brown.
\newblock Partitioning variables across register windows to reduce spill code
  in a low-power processor.
\newblock {\em IEEE Transaction on Computers}, 54(8):998--1012, 2005.

\bibitem{tallam2003a}
Sriraman Tallam and Rajiv Gupta.
\newblock Bitwidth aware global register allocation.
\newblock In {\em POPL}, pages 85--96, 2003.

\bibitem{woo2001a}
Seungdo Woo, Jungroin Yoon, and Jihong Kim.
\newblock Low-power instruction encoding techniques.
\newblock In {\em SOC Design Conference}, 2001.

\bibitem{xie2004a}
Fen Xie, Margaret Martonosi, and Sharad Malik.
\newblock Intraprogram dynamic voltage scaling: Bounding opportunities with
  analytic modeling.
\newblock {\em ACM Transactions on Architure and Code Optimization (TACO)},
  1(3):323--367, 2004.

\bibitem{zhang2002a}
Youtao Zhang and Rajiv Gupta.
\newblock Data compression transformations for dynamically allocated data
  structures.
\newblock In {\em 11th International Conference on Compiler Construction
  (CC'02), Lecture Notes in Computer Science}, volume 2304, pages 14--28,
  London, UK, 2002. Springer-Verlag.

\bibitem{zhuang2003a}
Xiaotong Zhuang, ChokSheak Lau, and Santosh Pande.
\newblock Storage assignment optimizations through variable coalescence for
  embedded processors.
\newblock In {\em LCTES '03: 2003 ACM SIGPLAN conference on Language, Compiler,
  and Tool for Embedded Systems}, pages 220--231, New York, NY, USA, 2003. ACM
  Press.

\end{thebibliography}

\end{document}